\documentclass{sig-alternate-05-2015}
\usepackage{amsmath,scalerel}
\usepackage{algorithm}
\usepackage{algorithmic}
\usepackage{tabularx, booktabs}
\usepackage{multirow}
\usepackage{slashbox}
\usepackage[keeplastbox]{flushend}
\usepackage[utf8]{inputenc}
\usepackage[english]{babel}
\usepackage{color}
\newcolumntype{C}{>{\centering\arraybackslash}p{10ex}}

\newcommand{\cmark}{\ding{51}}%
\newcommand{\xmark}{\ding{55}}%
\usepackage{pifont}

\newtheorem{definition}{Definition}

\begin{document}

\setcopyright{acmcopyright}
\conferenceinfo{SAC 2017}{3-7 April, 2017, Marrakesh, Morocco}

\title{Personalized Ranking for Context-Aware \\ Venue Suggestion}

\numberofauthors{3} 
\author{
\alignauthor Mohammad Aliannejadi \\
\affaddr{Universit{\`a} della Svizzera italiana (USI)} \\
\affaddr{Lugano, Switzerland}\\
\email{alianm@usi.ch}
\alignauthor Ida Mele \\
\affaddr{Universit{\`a} della Svizzera italiana (USI)} \\
\affaddr{Lugano, Switzerland} \\
\email{ida.mele@usi.ch}
\alignauthor Fabio Crestani \\
\affaddr{Universit{\`a} della Svizzera italiana (USI)} \\
\affaddr{Lugano, Switzerland} \\
\email{fabio.crestani@usi.ch}
 }

 
\maketitle
\begin{abstract}
Making personalized and context-aware suggestions of venues to the users is very crucial in venue recommendation. These suggestions are often based on matching the venues' features with the users' preferences, which can be collected from previously visited locations. In this paper we present a novel user-modeling approach which relies on a set of scoring functions for making personalized suggestions of venues based on venues content and reviews as well as users context.
Our experiments, conducted on the dataset of the TREC Contextual Suggestion Track, prove that our methodology outperforms state-of-the-art approaches by a significant margin.
\end{abstract}



\section{Introduction}
Nowadays, almost all mobile devices have Internet access which allows users to search for information wherever they are and whenever they need to. Users often rely on their mobile devices when they are looking for events to participate, activities to do, and interesting nearby venues to visit. 
In this paper we focus on venue suggestion, which consists of proposing a list of places that can be interesting for the user. This is an important task, since the traditional manual search for the best venue among the myriad of available ones may be time consuming and not easy to do, especially when the user is visiting a new city or, for example, wants to spontaneously plan the night out with friends.
Venue suggestions can be made by considering the preferences of the user, which are mined from the venues that the user has previously visited and can be further improved by the user's context (e.g., the user is alone or with family, she is on a business trip or on a romantic weekend).
 
In this paper we aim at making personalized suggestions by taking into account both the users' preferences and her context. Our approach assigns a score which depends on users' preferences, opinions, and context in order to rank the candidate suggestions.
Our model captures the user's preferences and understands her tastes by leveraging the venues' categories and the user's opinions. These are extracted from the online reviews often available in LBSNs.
The model is then enriched by adding the contextual information of a specific user  (e.g., season, group type).

The experiments on a TREC collection demonstrate that our approach performed very well compared to other state-of-the-art approaches~\cite{dean2015overview}.
\section{Related Work}\label{RelatedWork}
Recently, due to the availability of Internet access on mobile devices and on the fact that contextual information can be provided by the sensors of the mobile, researchers have been focusing their interest in context-aware suggestions for venues.
Compared to recommending news or products, the task of suggesting venues in a city raises further challenges, since it needs to consider not only the preferences of the users but also other constraints related to the context, such as the city, season, and people who accompany the user. 

Content-based approaches make suggestions for venues by simply  matching the venues' content (e.g., description and categories) with the user's preferences. Rikitianskii et al.~\cite{rikitianskii2014personalised} proposed to apply Part-of-Speech tagging to the venues' descriptions in order to get the most informative terms for a venue, which are then used to create positive and negative profiles. For each user, they trained a binary classifier using such profiles to rank the candidate suggestions. 

Review-based approaches aim to build enhanced user profiles using their reviews. Reviews provide a wealth of information that can be extracted to enable a system to deal with the data sparsity and cold-start problems.
Yang et al.~\cite{DBLP:conf/trec/Yang015} use reviews from Yelp to extract users' opinions. Given a pair \textit{(user, venue)}, they created positive and negative profiles for each pair by extracting data from all users' reviews. The list of suggestions is then ranked by using the similarity scores between all pairs of profiles.

In this paper, we propose a combination of content-based and review-based approaches. 
We use content to model users' interest and reviews to incorporate users' opinions in the model. We use also the context of a user since contextual information plays an important role in venue suggestion.
\section{User Modeling}\label{model}

\subsection{Frequency-based Score Component}\label{CategoryModel}
The first component is based on the frequencies of venue categories and taste tags. We first explain how to calculate the score for venue categories. The score for tags is calculated analogously.

Given a user $u$ and a her history of rated venues ($v_i$) $h_u = \{v_1, \dots, v_n\}$, each venue is assigned with a list of categories $C(v_i) = \{c_1, \dots, c_k\}$. We define the category profile of a user as follows:
\begin{definition}
A \textbf{Category Profile} is either positive or negative.
A Positive-category profile is a set of all distinct categories belonging to venues that a particular user has previously rated positively. A Negative-category profile is defined analogously for the venues that are rated negatively.
\end{definition}
We assign a user-level-normalized frequency value to each category in the positive/negative category profile. The user-level-normalized frequency for a positive/negative category profile is defined as follows:
\begin{definition}
A \textbf{User-level-Normalized Frequency} for an item (e.g., category) in a profile (e.g., positive-category profile) is calculated as follows: $\text{cf}^+(c_i) = \frac{count(c_i)}{\sum_{v_k \in h_u}\sum_{c_j \in C(v_k)} 1}$.
A user-level-normalized frequency for negative category profile, $cf^-$, is calculated analogously. 
\end{definition}

Given a user $u$ and a candidate venue $v$, the category-based similarity score $S_{cat}(u,v)$ between them is calculated as follows:
\begin{equation}
\label{eq:cat}
    S_{cat}(u,v) = \sum_{c_i \in C(v)}\text{cf}^+(c_i) - \text{cf}^-(c_i).
\end{equation}

We calculate the category similarity score from two sources of information, namely, Foursquare ($S_{cat}^{F}$) and Yelp ($S_{cat}^{Y}$).

\textbf{Venue Tags Score.}
We further enrich the category-based model using ``taste tags'' which are the most salient words extracted from the users' reviews.
We can leverage them to have a crisper description of the venues and improve our suggestions. We create positive and negative tag profiles for each user following Definition 1. Similar to the category scores, we assign a user-level-normalized frequency following Definition 2 to each tag, tf$(t)$, in the positive and negative tag profile. The tag similarity score is then calculated similar to Equation \ref{eq:cat}.

\subsection{Review-Based Score Component}\label{OpinionModel}
A further component uses the reviews to understand the motivation of the user behind a positive or negative rate. 
Indeed, modeling a user solely on venue's content is very general and does not allow to understand the reasons why the user enjoyed or disliked a venue.
Our intuition is that a user's opinion regarding an attraction could be learned based on the opinions of other users who gave the same or similar rating to the same attraction.

An alternative to binary classification would be a regression model, but we believe it is inappropriate since when users read online reviews, they make their minds by taking a binary decision (like/dislike). 
The binary classifier is trained using the reviews from the venues a particular user has visited before. 
We used the positive training samples which are extracted from the positive reviews of positive example suggestions,

Since the users' reviews contain lots of noise and off-topic terms, we calculated TF-IDF score as our feature vectors for training the classifier. As classifier we used Support Vector Machine (SVM) \cite{cortes1995support} with linear kernel and consider the value of the SVM's decision function as the score 
since it gives us an idea on how close and relevant a venue is to a user profile.

For each user we trained two SVM classifiers using reviews from Yelp and TripAdvisor. The corresponding scores are named $S_{rev}^{Y}$ and $S_{rev}^{T}$, respectively.

\subsection{Context-based Score Component}\label{ContextModel}
Contextual information is very important for improving the quality of venue suggestions. 
In this section, we propose two scores for measuring the similarity between the context of a user and the information about a place. Note that we are able to measure the contextual appropriateness of a venue to a given user only based on those contextual signals which are available on the LBSNs (i.e., the season, the trip, and the group type). 
Our basic idea is to compare the current user's context with the distribution check-ins of a particular venue over that context. We assume that the distribution of check-ins over a contextual signal reveals the level of the venues' appropriateness to that context. In the rest of this section we explain the score used for the season, a similar method is applied for the travel score.

\textbf{Season Score.} 
If we know in which season a user has been visiting a candidate venue, we can leverage the distribution of check-ins over seasons for a better ranking of venue suggestions. For those reviews which do not indicate the season, we assumed that most of the people leave reviews on LBSNs soon after they visit a place, and we can compute the distribution based on the reviews' timestamps. 

Let $S$ be the set of seasons and $s_u$ be the season a particular user visited a place, $SP(s, v)$ is a function returning the number of check-ins by other users for venue $v$ in season $s$. Hence, we define the season score for user $u$ visiting $v$ as:
\begin{equation}
    \label{eq:season}
    S_{cxt}^{season}(u,v) = SP(s_u, v) - \frac{\sum_{s_i \in S, s_i \neq s_u}SP(s_i, v)}{\sum_{s_i \in S, s_i \neq s_u}1},
\end{equation}
where $SP(s_u, v)$ is the number of check-ins for the venue $v$ in the same season of the user $u$ and $\sum_{s_i \in S, s_i \neq s_u}1$, is the number of seasons other than the season when the user data was recorded.
This score effectively detects if a venue is appropriate for a specific season by dividing the four seasons into two buckets: one is the user's current season and the other one is given by the other seasons for which the average is  computed. 

\textbf{Travel Score.} We can assume two more dimensions in user's context which can be leveraged to enhance the personalized ranking of venues. These two types of information are \textit{Trip Type} and \textit{Group Type}. Trip type indicates whether the user is visiting a venue for business or leisure, while group type defines the group that is accompanying the user in her trip (e.g., family, friends, etc.) We looked into the information available on some LBSNs to find the best possible match between such contextual information and the information about places. Some LBSNs track and report distribution of traveler types who visit a particular place. Therefore, we map these two contextual dimensions onto the available information from TripAdvisor. The score $S_{cxt}^{travel}$ is calculated similar to Equation \ref{eq:season}.
\section{Experimental Results}\label{exp}

\textbf{Dataset.} 
We used the dataset of TREC Contextual Suggestion Track 2015.
More in details, given a set of example venues as user's preferences and some contextual signals, the task consists in returning a ranked list of candidate venues which fit the user's profile and context.

\textbf{Evaluation Metrics.}
We evaluate the performance of our proposed model by reporting P@5 (Precision at 5) and MRR (Mean Reciprocal Rank). Our model uses \textbf{Ca}tegory, \textbf{T}ags, \textbf{Re}views, and \textbf{C}onte\textbf{x}t, so we call it \textbf{CaTReCx}.

\textbf{Baselines.}
We compared our method with the three top-ranked participants in the TREC Contextual Suggestion Track 2015 \cite{dean2015overview}. 
Our first baseline is the best performing run (BASE1)~\cite{alian2015}  which uses four scores (reviews from Yelp, categories from Yelp and TripAdvisor, and keywords from Foursquare). It ranks venues based on the linear combination of these scores. 
The second baseline is the second best run 
(BASE2)~\cite{mccreadie2015university} that utilizes factorization machines for venue recommendation. The instances that are fed into the factorization machine are composed of three blocks representing user, context, and venue features. It uses Foursquare as its source of information. The third best run 
(BASE3)~\cite{DBLP:conf/trec/Yang015} creates positive and negative profiles for each user and adds to them all the reviews of similar users from Yelp. It creates positive and negative profiles for venues. The venues are then ranked by linearly combining the similarity scores of all profile pairs. Finally, the ranked list of venues is modified by applying a number of contextual filters.

\textbf{Results.}
We ranked the venues considering all the aforementioned scores as features for LambdaMART learning-to-rank technique. We conducted our experiments using a 5-fold cross validation across the training data. Table \ref{tb:res1} shows the performance of our model as well as of the baselines. 
Experimental results demonstrate that our system outperforms all the three baselines by a significant margin when it uses all the three sources of information. Note that in order to perform a fair comparison between our work and the baselines, we also report our system's performance using only the same source of information used by the respective baseline. More in details, CaTReCx achieves a 5.34\% improvement in terms of P@5 and a 3.93\% improvement in terms of MRR over BASE1. Our approach also exhibits an improvement over BASE2 using only the data from Foursquare (F) in terms of P@5. Moreover, it beats BASE3 in terms of both P@5 and MRR by a large margin. For completeness, we report the median performance of all participants of TREC. The performance of our methodology compared to the TREC median performance proves the effectiveness of our model. 
In particular, since our approach combines multimodal information from multiple LBSNs, it can significantly improve the precision of venue suggestions. It is worth noting that we tried different classifier and regression algorithms for the review-based score component. However, since the SVM classifier with a linear kernel exhibited a much better performance than the other models we do not report the results of the others.
In fact, as we discussed in our previous work~\cite{alianAIRS2016} SVM is a perfect match for this classification problem since the number of positive training samples are much more than the negative samples. Most classifiers tend to correctly classify the class with more training examples, while SVM is not affected by the relative size of the classes.

\begin{table}[ht]
    \centering
    \caption{Performance comparison with  other TREC participants. \textit{Y} stands for  Yelp,  \textit{T} for TripAdvisor and \textit{F} for Foursquare.}
    \begin{tabular}{l c c c c c}
     \toprule
      & Y & T & F & P@5 & MRR \\ 
     \midrule
      CaTReCx & \cmark & \cmark & \cmark & \boldmath$0.6171$ & \boldmath$0.7695$\\ 
      BASE1 & \cmark & \cmark & \cmark & $0.5858$ &  $0.7404$\\ 
      \midrule
      
      CaTReCx & \xmark & \xmark & \cmark & \boldmath$0.5782$ & $0.7188$\\ 
      BASE2 & \xmark & \xmark & \cmark & $0.5706$ & \boldmath$0.7190$ \\
      \midrule
      
      CaTReCx & \cmark & \xmark & \xmark & \boldmath$0.6095$ & \boldmath$0.7453$\\ 
      BASE3 & \cmark & \xmark & \xmark & $0.5583$ & $0.6815$ \\
    \hline
     TREC Median & - & - & - & $0.5090$ & $0.6716$ \\ 
     \bottomrule
    \end{tabular}
    \label{tb:res1}
\end{table}

\section{Conclusions and Future Work}\label{Conclusions}
In this paper we described a personalized ranking model for context-aware venue suggestion. Our model aimed to capture various types of information from multiple sources which can be important to a user for visiting a venue. 
The experimental results on the TREC Contextual Suggestion Track dataset demonstrated our system effectiveness compared to the state of the art.

As future work, it would be interesting to define a generative probabilistic model which predicts user tags for a new venue by modeling empirically the mapping between the tags the user selected for a venue and its content.

\section{Acknowledgments}
This work was partially supported by the RelMobIR project of the Swiss National Science Foundation (SNSF).

\bibliographystyle{abbrv}
\bibliography{main} 

\end{document}